\documentstyle[12pt]{article}
\newcommand{\slq}{\raise.15ex\hbox{$/$}\kern-.57em\hbox{$q$}}
\newcommand{\slp}{\raise.15ex\hbox{$/$}\kern-.57em\hbox{$p$}}
\newcommand{\be}{\begin{equation}}
\newcommand{\ee}{\end{equation}}
\newcommand{\bear}{\begin{eqnarray}}
\newcommand{\ear}{\end{eqnarray}}

\newcommand{\gsim}{\stackrel{\scriptstyle >}{\sim}}
\newcommand{\lsim}{\stackrel{\scriptstyle <}{\sim}}

\date{}
\begin{document}
\begin{titlepage}
\begin{flushright}
HD--THEP--96--2
\end{flushright}
\quad\\
\vspace{1.8cm}
\begin{center}
{\bf\Large SYSTEMATIC RESUMMED PERTURBATION THEORY}\\
\vspace{1cm}
Christof Wetterich\\
\bigskip
Institut  f\"ur Theoretische Physik\\
Universit\"at Heidelberg\\
Philosophenweg 16, D-69120 Heidelberg\\
\vspace{1cm}
\end{center}
\begin{abstract}
A systematic loop expansion is formulated in terms of full
propagators and vertices. It is based on an expansion of the
general solution of an exact non-perturbative flow equation.
\end{abstract}
\end{titlepage}

One-particle irreducible vertices are the basic building
blocks for the computation of cross sections and decay rates
in particle physics or for the critical equation of state
in statistical physics. They are often computed in a perturbative
or loop expansion as functions of renormalized coupling constants.
In some circumstances one would like to use instead of constant
couplings the full momentum-dependent vertices and propagators
in the loop integrals. As an example we mention the scale
ambiguity: The renormalized couplings used in perturbation theory
are defined by $n$-point functions evaluated at a given scale $\mu$
for the momenta. Renormalization group equations describe the scale
dependence of the corresponding renormalized coupling $g(\mu)$.
A one-loop calculation for some different
$n$-point function corresponds to
a certain power of $g(\mu)$ and one always encounters a scale
ambiguity: At what scale should the coupling be taken? The
difference between the choice of two distinct scales is of
higher loop order and cannot be settled within the one-loop
calculation. On the other hand, one often has knowledge about
the momentum dependence of the full vertex by which $g(\mu)$
is defined, and one may be tempted to use this for a more
accurate determination already of the one-loop graph. If one could
replace $g(\mu)$ in the loop integral by the full momentum-dependent
vertex, the scale ambiguity would disappear. There would be
no need to select a particular renormalization scale $\mu$ since
one integrates over the momenta appearing in the vertex. The
inherent difficulty in such an approach concerns the compatibility
with higher loop orders. Since the use of momentum-dependent
vertices corresponds to a resummation of higher orders of $g(\mu)$,
one must avoid a double counting of these effects once the two-loop
correction is included. The question arises of how to formulate a
systematic expansion where only full vertices and propagators
instead of the (renormalized) classical ones appear in every order
of the expansion. Applications of such an expansion would be much
wider than the mentioned scale ambiguity problem. We only mention
here the use of solutions of gap equations in loop integrals for
other $n$-point functions. There appears again a double counting
problem in two-loop order which has to be treated systematically.

We propose in this letter a way of avoiding such
double counting by developing a systematic resummed perturbation
theory formulated directly in terms of renormalized $n$-point
functions. It is based on the formal solution of an exact flow
equation which describes the change of the effective action $\Gamma_k$
with the infrared cutoff scale $k$ in terms of a renormalization-group
improved one-loop expression \cite{FE}. The exact flow equation
is closely related to the Wilsonian approach to the renormalization
group equations \cite{RGE}. At the end of this letter we will sketch
several possible applications of systematic resummed perturbation
theory (SRPT)
which go beyond the range of problems accessible within standard
perturbation theory.

We start from the exact non-perturbative flow equation\footnote{We
restrict the discussion in this letter to bosons. The generalization
to fermions is straightforward, with a supertrace in (\ref{1}).}
$(t=\ln\frac{k}{\Lambda})$
\be\label{1}
\partial_t\Gamma_k[\varphi]=\frac{1}{2}\ Tr\ \left\{
\partial_t {\cal R}_k\ (\Gamma_k^{(2)}[\varphi]+{\cal R}_k)
^{-1}\right\}\ee
It is obtained by adding to the classical action a quadratic infrared
cutoff
\be\label{2}
\Delta_kS=\frac{1}{2}\ Tr\ \left\{{\cal R}_k\ (\bar\chi\otimes
\chi)\right\}\ee
and differentiating the functional integral which defines the
$k$-dependent effective action $\Gamma_k$ with respect to $k$
\cite{FE}. Here the trace includes a summation over internal
indices as well as a momentum integration,
$Tr=\int\frac{d^dq}{(2\pi)^d}\sum_a$, and
$\chi$ stands for the fluctuations
$\chi_a(q)$ over which the
functional integral is performed. The flow equation (\ref{1})
involves the exact field-dependent inverse propagator
\be\label{3}
(\Gamma_k^{(2)})_{ab}(q,q')=\frac{\delta^2\Gamma_k}{\delta
\varphi_a(q)\delta\bar\varphi_b(q')}\ee
(We use here a language with matrices $\Gamma_k^{(2)}, {\cal R}_k$
and $(\bar\chi\otimes\chi)_{ba}(q',q)=\bar\chi_b(q')\chi_a(q)$.)
The infrared cutoff ${\cal R}_k$ is typically (but not
necessarily) diagonal in momentum space and should vanish for $k\to 0$.
An example is
\be\label{4}
({\cal R}_k)_{ab}(q,q')=Z_kq^2(e^{q^2/k^2}-1)^{-1}(2\pi)^d
\delta(q-q')\delta_{ab}\ee
The ``average action'' $\Gamma_k$ is a functional of
``classical'' fields $\varphi$. For $k\to0$ it becomes the usual
effective action, i.e. the generating functional for the 1PI Green
functions. Our aim is a suitable loop expansion
of the general solution of the flow equation (\ref{1})
for $\Gamma[\varphi]=
\Gamma_{k\to0}[\varphi]$. Indeed, we may use the flow equation
for an implicit definition of the theory, with all short-distance
physics and regularization encoded in the ``initial value''
$\Gamma_\Lambda[\varphi]$ specified at some high momentum scale
$\Lambda$.

Standard perturbation theory can easily be recoverd from an iterative
solution of the flow equation (\ref{1}). Starting from the leading
or ``classical'' contribution $\Gamma_{k(0)}\equiv\Gamma_\Lambda$
one may insert this instead of $\Gamma_k$ on the r.h.s. of (\ref{1}).
Performing the $t$-integration generates the one-loop
contribution
\be\label{5}
\Gamma_k-\Gamma_\Lambda=\frac{1}{2} Tr\left\{
\ln (\Gamma_\Lambda^{(2)}+{\cal R}_k)-\ln(\Gamma_\Lambda^{(2)}+{\cal
R}_\Lambda)\right\},\ee
where we remind that ${\cal R}_k\to0$ for $k\to 0$.
We observe that the momentum integration on the r.h.s. of (\ref{5})
is regularized in the ultraviolet through subtraction
of $\ln(\Gamma_\Lambda^{(2)}+{\cal R}_\Lambda)$. This is a type
of implicit Pauli-Villars regularization with the heavy mass
term replaced by a momentum-dependent piece ${\cal R}_\Lambda$ in the
inverse propagator. With suitable chirally invariant ${\cal R}_\Lambda$
\cite{CC} this can be used for a regularization of models with chiral
fermions. Also gauge theories can be regularized in this way, but
care is needed since $\Gamma_\Lambda$ has to obey identities
reflecting the gauge invariance \cite{GI}. Going further, the
two-loop contribution obtains by inserting the one-loop expression
for $\Gamma_k^{(2)}$ as obtained from eq. (\ref{5})
into the r.h.s. of (\ref{1}). Only the
classical inverse propagator $\Gamma_\Lambda^{(2)}$ and its
functional derivatives appear in the nested expressions. They are
independent of $k$ and the integration of the approximated flow equation
is straightforward.

It is the purpose of this letter to replace the perturbative
iteration sketched above by a new one which involves the full
propagators and vertices instead of the classical ones. We start
again with the lowest order term
\be\label{6}
\Gamma_{k(0)}[\varphi]=\Gamma_\Lambda[\varphi]\ee
where $\Lambda$ is now some conveniently chosen scale (not
necessarily the ultraviolet cutoff). In the next step we write equation
(\ref{1}) in the form
\be\label{7}
\partial_t\Gamma_k=\frac{1}{2}\ Tr\ \partial_t\ln(\Gamma_k^{(2)}
+{\cal R}_k)
-\frac{1}{2}\ Tr\ \left\{\partial_t\Gamma_k^{(2)}(\Gamma_k
^{(2)}+{\cal R}_k)^{-1}\right\}\ee
Here $\partial_t\Gamma_k^{(2)}$ can be inferred by taking
the second functional derivative of eq. (\ref{1})  with respect
to the fields $\varphi$. Equation (\ref{7})
can be taken as the starting point of a systematic loop expansion
by counting any $t$-derivative acting only on $\Gamma_k$
or its functional derivatives as an
additional order in the number of loops. From (\ref{1}) it is
obvious that any such derivative involves indeed a new momentum
loop. It will become clear below that in case of weak
interactions it also
involves a higher power in the coupling constants. The (modified)
one-loop contribution $\Gamma_{k(1)}'$ can now be defined by
\be\label{8}
\Gamma_k=\Gamma_{k(0)}+\Gamma_{k(1)}'+\Gamma^{(R)}_{k(1)}\ee
with
\be\label{9}
\Gamma_{k(1)}'=\frac{1}{2}\ Tr\ \left\{\ln(\Gamma_k^{(2)}+{\cal R}_k)-
\ln(\Gamma_k^{(2)}+{\cal R}_\Lambda)\right\}\ee
In contrast to eq. (\ref{5}) the r.h.s. involves now the full
(field-dependent) inverse propagator $\Gamma_k^{(2)}$, and, by
performing suitable functional derivatives, the full proper vertices.
Putting $k=0$ the resummed one-loop expression (\ref{9})
ressembles a Schwinger-Dyson \cite{SD}
or gap equation, but in contrast to
those only full vertices appear. For example, for $k=0$ the one-loop
contribution to the inverse propagator obtains by taking the second
functional derivative of eq. (\ref{9})
\bear\label{10}
\left({\Gamma'}^{(2)}_{(1)}\right)_{ab}(q,q')&=&\frac{1}{2}\ Tr
\left\{\left(\Gamma^{(2)}\right)^{-1}
\frac{\delta^2\Gamma^{(2)}}{\delta\varphi_a(q)
\delta\bar\varphi_b(q')}\right\}\nonumber\\
&&-\frac{1}{2}\ Tr\ \left\{\left(\Gamma^{(2)}\right)^{-1}
\frac{\delta\Gamma^{(2)}}{\delta\varphi_a(q)}\left(\Gamma^{(2)}\right)
^{-1}\frac{\delta\Gamma^{(2)}}{\delta\bar\varphi_b(q')}\right\}
\nonumber\\
&&- {\rm regulator\ terms}\ear
and involves the proper three- and four-point vertices. Adding the
lowest order piece $\Gamma_{(0)}^{(2)}$ and approximating the vertices
by their lowest order expressions, eq. (\ref{10})
reduces to the standard
gap equation for the propagator in a regularized form. Assuming that
this is solved (for example numerically by an iterative procedure)
we see that the resummed one-loop expression (\ref{10}) involves
already arbitrarily high powers in the coupling constant, and,
in particular, contains part of the perturbative two-loop contribution.
The remaining part of the perturbative two-loop contribution
must appear in the resummed two-loop contribution that we will
discuss next.

The remaining piece beyond resummed one-loop order $\Gamma_{k(1)}
^{(R)}$ obeys the flow equation
\bear\label{11}
&&\partial_t\Gamma^{(R)}_{k(1)}=-\frac{1}{2}\ Tr\ \left\{\partial_t
\Gamma_k^{(2)}\left[\left(\Gamma_k^{(2)}+{\cal R}_k\right)^{-1}
-\left(\Gamma_k^{(2)}+{\cal R}_\Lambda\right)^{-1}\right]\right\}
\nonumber\\
&&=\frac{1}{4}\left[\left(\Gamma_k^{(2)}+
{\cal R}_k\right)^{-1}-\left(\Gamma_k^{(2)}
+{\cal R}_\Lambda\right)^{-1}\right]_{\sigma
\tau}\frac{\delta^4\Gamma_k}{\delta
\varphi_\tau\delta\bar\varphi_\sigma
\delta\varphi_\alpha\delta\bar\varphi_\beta}(\Gamma_k^{(2)}+
{\cal R}_k)^{-1}_{\beta\gamma}\nonumber\\
&&\qquad(\partial_t{\cal R}_k)_{\gamma\delta}
(\Gamma_k^{(2)}+{\cal R}_k)^{-1}_{\delta\alpha}
\nonumber\\
&&-\frac{1}{2}\left[\left(\Gamma_k^{(2)}+{\cal R}_k\right)^{-1}-
\left(\Gamma_k^{(2)}+{\cal R}_\Lambda\right)^{-1}\right]_{\sigma\tau}
\frac{\delta^3\Gamma_k}{\delta\varphi_\tau\delta\varphi_\alpha
\delta\bar\varphi_\beta}\left(\Gamma_k^{(2)}+{\cal R}_k\right)^{-1}
_{\beta\gamma}\nonumber\\
&&\qquad\frac{\delta^3\Gamma_k}{\delta\bar
\varphi_\sigma\delta\varphi_\gamma
\delta\bar\varphi_\delta}\left(\Gamma_k^{(2)}+{\cal R}_k\right)^{-1}
_{\delta\epsilon}(\partial_t{\cal R}_k)_{\epsilon\eta}\left(\Gamma_k
^{(2)}+{\cal R}_k\right)^{-1}_{\eta\alpha}\ear
where $\alpha,\beta,...$ etc combine internal indices and momentum
labels. Using the symmetries of (\ref{11}) it is straightforward
to extract the resummed two-loop contribution
\be\label{12}
\Gamma_{k(1)}^{(R)}=\Gamma_{k(2)}'+\Gamma^{(R)}_{k(2)}\ee
by writing (\ref{11}) as a total $t$-derivative
plus terms where $\partial_t$ acts only on $\Gamma_k$ and its
functional derivatives. One finds
\bear\label{13}
\Gamma_{k(2)}'&=&-\frac{1}{8}\ \frac{\delta^4\Gamma_k}{\delta
\varphi_\tau\delta\bar\varphi_\sigma\delta\varphi_\alpha
\delta\bar\varphi_\beta}G_{\sigma\tau}G_{\beta\alpha}\nonumber\\
&&+\frac{1}{6}\ \frac{\delta^3\Gamma_k}{\delta\varphi_\tau\delta\varphi
_\alpha\delta\bar\varphi_\beta}\ \frac{\delta^3\Gamma_k}
{\delta\bar\varphi_\sigma\delta\varphi
_\gamma\delta\bar\varphi_\delta}\
G_{\sigma\tau}G_{\beta\gamma}\left(G_{\delta\alpha}+
\frac{3}{2}\left(\Gamma_k^{(2)}+{\cal R}_\Lambda\right)
^{-1}_{\delta\alpha}\right)\ear
with regularized propagator
\be\label{14}
G_{\sigma\tau}=\left(\Gamma_k^{(2)}+{\cal R}_k\right)^{-1}_{\sigma
\tau}-\left(\Gamma_k^{(2)}+{\cal R}_\Lambda\right)
^{-1}_{\sigma\tau}\ee
At this point it becomes obvious how the iterative construction of
higher resummed loop terms proceeds: The flow equation for
$\Gamma_{k(2)}^{(R)}$ can be written in the form
\be\label{15}
\partial_t\Gamma^R_{k(2)}=(\tilde\partial_t-\partial_t)\Gamma_{k(2)}'\ee
with $\tilde\partial_t$ acting only on ${\cal R}_k\
\left(\tilde\partial_t=
(\partial_t{\cal R}_k)\frac{\partial}{\partial {\cal R}_k},\
\tilde\partial_t\Gamma_{k(2)}'=\partial_t\Gamma^{(R)}_{k(1)}\right)$.
It involves $t$-derivatives acting on $\Gamma_k$ and its
functional derivatives with respect to the fields. Using (\ref{1}),
they can be expressed in terms of $\partial_t{\cal R}_k$ in one higher
loop
order. The resummed three-loop contribution obeys now $\tilde\partial
_t\Gamma_{k(3)}'=\partial_t\Gamma_{k(2)}^{(R)}$ and so on.

Comparing the resummed two-loop contribution (\ref{13}) with the
standard two-loop contribution, we find the same type of graphs
(cf. fig. 1).
Only the weight factors have changed - they are now $-\frac{1}{8}$ and
$\frac{1}{6}$ instead of $\frac{1}{8}$ and
$-\frac{1}{12}$ in standard perturbation theory. And, of course, the
graphs involve now full propagators and vertices instead of the classical
ones. The difference in the weight factors is easily understood: The
resummed one-loop contribution $\Gamma_{(1)}'$ also contains terms
with the same structure as the two-loop contribution, once we expand
eq. (\ref{9}) iteratively.
In fact, the functional form of $\Gamma_k$ is not known
exactly and one has to use a truncation $\Gamma_k^{(tr)}$
on the r.h.s. of eqs. (\ref{9}) and (\ref{13}). We expand
$\Gamma_{k(1)}'$ in powers of the difference $D_k=\Gamma_k
-\Gamma_k^{(tr)}$
\be\label{16}
\Gamma_{k(1)}'=\Gamma_{k(1)}+\Delta\Gamma_{k(2)}+...\ee
with truncated one-loop contribution
\be\label{17}
\Gamma_{(1)}=\frac{1}{2}\ {\rm Tr}\ln\Gamma^{(tr)(2)}\ee
(We use here $k=0$ and omit the regulator terms.)
The second piece $\Delta\Gamma_{k(2)}=\frac{1}{2}\ Tr\
G^{(tr)}D^{(2)}_{k(1)}$ obtains by inserting the lowest order for
$D_k$
\bear\label{18}
D_{k(1)}&=&\frac{1}{2}Tr\left\{\ln\left(\Gamma_k^{(tr)(2)}
+{\cal R}_k\right)-\ln\left(\Gamma_k^{(tr)(2)}+{\cal R}
_\Lambda\right)\right\}
\nonumber\\
&&-\left(\Gamma_k^{(tr)}-\Gamma_\Lambda\right)\ear
and is counted as a two-loop contribution
\bear\label{19}
\Delta\Gamma_{k(2)}&=&\Delta\Gamma_{k(2)}^{(ct)}+
\frac{1}{4}\frac{\delta^4\Gamma_k^{(tr)}}
{\delta\varphi_\tau\delta\bar\varphi_\sigma\delta\varphi_\alpha
\delta\bar\varphi_\beta}G^{(tr)}_{\sigma\tau}G^{(tr)}_{\beta\alpha}
\\
&&-\frac{1}{4}\frac{\delta^3\Gamma_k^{(tr)}}{\delta\varphi_
\tau\delta\varphi_\alpha\delta\bar\varphi_\beta}\
\frac{\delta^3\Gamma_k^{(tr)}}{\delta\bar\varphi_\sigma\delta\varphi
_\gamma\delta\bar\varphi_\delta}G_{\sigma\tau}^{(tr)}
G_{\beta\gamma}^{(tr)}
\left(G_{\delta\alpha}^{(tr)}+2\left(\Gamma_k^{(tr)(2)}+
{\cal R}_\Lambda\right)^{-1}_{\delta\alpha}\right)\nonumber
\ear
\be\label{20}
\Delta\Gamma_{k(2)}^{(ct)}=\frac{1}{2}\ Tr\left\{G^{(tr)}
\left(\Gamma_\Lambda^{(2)}-\Gamma_k^{(tr)(2)}\right)\right\}\ee
We notice that $\Delta\Gamma^{(ct)}_{k(2)}$ has the structure
of a counter term insertion in a one-loop integral whereas the last
two terms in eq. (\ref{19}) correspond to the graphs in fig. 1 with
weight factors $\frac{1}{4}$ and $-\frac{1}{4}$. Combining
$\Delta\Gamma_{(2)}$ with the lowest order of $\Gamma_{(2)}'$
(\ref{13}) one finally obtains the truncated two-loop contribution
($k=0$ and omitting regulator terms)
\bear\label{21}
\Gamma_{(2)}&=&\frac{1}{8}\left(\Gamma^{(tr)(2)}\right)^{-1}
_{\sigma\tau}\frac{\delta^4\Gamma^{(tr)}}{\delta\varphi_\tau
\delta\bar\varphi_\sigma\delta\varphi_\alpha\delta\bar\varphi_\beta}
\left(\Gamma^{(tr)(2)}\right)^{-1}_{\beta\alpha}\\
&&-\frac{1}{12} \left(\Gamma^{(tr)(2)}\right)^{-1}_{\sigma\tau}
\frac{\delta^3\Gamma^{(tr)}}{\delta\varphi_\tau\delta\varphi_\alpha
\delta\bar\varphi_\beta}\left(\Gamma^{(tr)(2)}\right)^{-1}_{\beta
\gamma}\frac{\delta^3\Gamma^{(tr)}}{\delta\bar
\varphi_\sigma\delta\varphi_\gamma\delta\bar\varphi_\delta}
\left(\Gamma^{(tr)(2)}\right)^{-1}_{\delta\alpha}+\Delta
\Gamma_{(2)}^{(ct)}\nonumber\ear

For the simplest ansatz $\Gamma^{(tr)}=\Gamma_\Lambda$ one recovers
standard unrenormalized perturbation theory.
We note that renormalized
perturbation theory in a given scheme can also be recovered
directly in our language as a special case:
It corresponds to the truncation where $\Gamma^{(tr)}$ consists
only of the $n$-point functions appearing in the classical
action, with coefficients given by
renormalized couplings as determined by the scheme-dependent
renormalization conditions. Consider the example
of the $\varphi^4$-theory with renormalized quartic coupling
$\lambda(\mu)$ defined by the four-point
function at symmetric momenta.
In this scheme the contribution to the
four-point function contained in $D$ is
$\sim\Delta\lambda_4(q_1,q_2,
q_3,q_4)\varphi(q_1)\varphi(q_2)\varphi(q_3)\varphi(q_4)$ with
$\Delta\lambda_4=0$ for symmetric momenta with scale $\mu$.
The scheme  and scale dependence of the
perturbative results appears in this language through the
dependence of $D$ on the renormalization condition and $\mu$.
In close analogy, the one-loop result of truncated SRPT will
always exhibit a ``truncation dependence'', but not necessarily
the usual scale dependence.

Systematic resummed perturbation theory is particularly
convenient for a computation of ultraviolet finite $n$-point
functions (as the $\varphi^6$ coupling) or differences of
$n$-point functions at different momenta (e.g. $\Delta\lambda_4$).
For $\Lambda\to\infty$ all dependence on the effective ultraviolet
cutoff $\Lambda$ is absorbed in the renormalized couplings. In
this limit the regulator terms vanish.
This is very useful in the case of QCD or other gauge
theories where calculations with an explicit regulator function
${\cal R}_\Lambda$ are quite cumbersome\footnote{One can also
combine SRPT with dimensional regularization such that the limit
$\Lambda\to\infty$ can be taken even for UV-divergent quantities.}.
(Of course, ultraviolet finiteness requires here that appropriate
Slavnov-Taylor or background field identities \cite{GI}
are respected by $\Gamma^{(tr)}$.) We note that all Green functions
can be made ultraviolet finite by appropriate subtractions. For the
two-point function a typical infrared-finite quantity is
$\lambda_2(q)-\lambda_2(0)
-\frac{\partial\lambda_2}{\partial q^2}_{|q^2=\mu^2}q^2$ with
$\Gamma^{(2)}=\lambda_2(q)(2\pi)^d\delta(q-q')$ for constant
vacuum fields.

The necessary cancellation of the cutoff$(\Lambda)$-dependence
for ultraviolet finite quantities has some subtleties once the
truncation goes beyond renormalized perturbation theory. In
this case a given
loop order in SRPT involves for the higher $n$-point functions
also diagrams not present in standard perturbation theory. For
example, the one-loop order for the four-point function obtains by
suitable functional differentiation of eq. (\ref{9}) or (\ref{17})
and involves up
to six-point functions.\footnote{The relation to standard perturbation
theory follows from an iterative solution for
the higher $n$-point functions.}
We have depicted in fig. 2 the contribution from
the $\varphi^6$ coupling in a typical scalar theory. In the limit
$\Lambda\to\infty$ one encounters not only the usual ultraviolet
divergences. Higher order couplings present in $\Gamma^{(tr)}$ and
absent in the classical action can induce additional divergences,
as for example a possible
divergence $\sim\Lambda^2$ from the graph of fig. 2
for the four-dimensional
scalar theory. This divergence is an
artefact of the expansion and cancelled by two-loop
contributions of the type depicted in fig. 3.
The exact definition of truncated SRPT and therefore the
structure of divergences in a given loop order
is specified by the ansatz for $\Gamma_k^{(tr)}$.\footnote{The optimum
choice can be adapted to the quantity one wants to compute.}
A good guide is certainly to avoid ultraviolet divergences
in one-loop order for quantities which must be ultraviolet finite.
This may require a tuning of the ``higher order couplings'' (e.g. the
$\varphi^6$ coupling $\lambda_6$ or $\Delta\lambda_4$) used in the
ansatz for $\Gamma^{(tr)}$. Such a tuning is no accident since
the values of the higher order couplings in a given model are
fixed by an infrared fixed-point behaviour. They are no free parameters
but rather computable in terms of the renormalized couplings.
The requirement of vanishing ultraviolet divergences can be used
as a constraint for the higher order couplings taken into account
in $\Gamma_k^{(tr)}$.

The aim of the proposed systematic resummed perturbation theory
is not only the computation of Green functions
in a given order in the coupling constant. For this purpose
standard perturbation theory is sometimes more direct and efficient.
The systematic
character of SRPT guarantees that in every loop order the
standard perturbative contributions are all included. Beyond that,
however, a given loop order in SRPT also contains partial
higher-loop effects of standard perturbation theory. The convergence
of SRPT is not necessarily related to the existence of a small
coupling constant. From eq. (\ref{7})
we learn that the total size of all higher-loop contributions
as compared to the one-loop contribution is controled by the
relative size of the piece arising from $\partial_t\Gamma_k^{(2)}$ as
compared to $\partial_t(\Gamma_k^{(2)}+R_k)$. In particular, if
$\Gamma_k$ changes only slowly with $k$ - for example in case of
an effective physical IR-cutoff - one expects SRPT to converge
rather well.

The virtues of SRPT particularly pertain
to situations at the borderline of validity of standard
perturbation theory and the possible combination with non-perturbative
approaches. Among the possible useful applications we mention
the combination of SRPT with gap equations for the
mass term. Gap equations are frequently used for situations
where the renormalized mass terms differ substantially from the
bare mass term (even after subtraction of counterterms), as, for
example, in case of spontaneous symmetry breaking or high
temperature field theory. SRPT allows for the use of solutions of
gap equations within a systematic loop expansion. This becomes
possible since $\Gamma_k^{(tr)(2)}$ instead of
$\Gamma_\Lambda^{(2)}$ appears
systematically in all graphs. In addition, SRPT may also be used as
a starting point for the gap equation
which determines the mass term in $\Gamma^{(tr)}$
(cf. eq. (\ref{10})).
It has the advantage that no short-distance couplings (bare couplings)
enter this equation. (The latter contrasts with the Schwinger-Dyson
equations which are often used as a starting point for gap equations.)

A second issue concerns the possibility of a systematic split of loop
graphs into high and low momentum contributions. This may be
advantageous when the physics of the low momentum modes differs
qualitatively from the one for the
high momentum modes, as in QCD. In fact,
the successful QCD sum rules are employed in this spirit, even though
not fully systematic so far.
A given $n$-loop contribution of SRPT can always
be split
\be\label{22}
 \Gamma_{(n)}=\Gamma_{k_c(n)}+\left(\Gamma_{(n)}-\Gamma_{k_c(n)}
\right)\ee
where $\Gamma_{k_c(n)}$ contains an explicit IR-cutoff
${\cal R}_{k_c}$
and only fluctuations with momenta $q^2\gsim k^2_c$ are effectively
included in this piece. The second piece, $\Gamma_{(n)}-
\Gamma_{k_c(n)}$, includes the remaining effects of modes with
$q^2\lsim k_c^2$. For this piece $k_c$ acts as
an effective ultraviolet cutoff which guarantees that only the low
momentum modes are counted. For a demonstration, let us consider
the one-loop contribution (\ref{9}) with $\Lambda\equiv k_c,\ k=0$.
This assumes that one has already computed the ``perturbative part''
$\Gamma_{k_c}$ which acts now as the classical part
$\Gamma_{(0)}=\Gamma_\Lambda=\Gamma_{k_c}=\Gamma^{(tr)}$
for the remaining
calculation. The ``non-perturbative part''
\be\label{23}
\Gamma_{(1)}^{(np)}\equiv \Gamma_{(1)}-\Gamma_{k_c(1)}=\frac{1}{2}
\ Tr \left\{\ln\Gamma^{(2)}-\ln\left(\Gamma^{(2)}+{\cal R}
_{k_c}\right)
\right\}\ee
encodes the infrared physics and exhibits the UV-cutoff
$k_c$. Functional derivatives of this quantity (for example
with respect to heavy quark fields in a QCD calculation) lead to
the appearance of momentum integrals over propagators (for example
for gluons). These momentum integrals are UV-regularized and can
be associated with expectation values of
corresponding regularized composite operators.
In a QCD computation this involves regularized operators for gluon
and quark condensates. which can be parametrized
phenomenologically. For a systematic
formulation of these concepts it
is obviously crucial that the full propagator
$\left(\Gamma^{(2)}\right)^{-1}$ and not the classical propagator
appears in the relevant momentum integrals.

Finally, SRPT could be used in connection with approximative
solutions of the flow equation (\ref{1}) in the following way:
Eq. (\ref{1}) is a functional differential equation which cannot
be reduced to a closed system for a finite number of couplings.
For example, the beta function for the four-point vertex (the fourth
functional derivative of the r.h.s. of eq. (\ref{1})) involves
not only two-, three- and four-point functions, but also up to
six-point functions. Approximate solutions to the flow equation
often proceed by truncation. For example, contributions involving
the five- and six-point function could be neglected. As an
alternative, these higher $n$-point functions could be evaluated
by SRPT. (Closely related ideas have already been tested
successfully for
four-dimensional scalar theories \cite{PW}.) We observe that the
momentum integrals relevant for the higher $n$-point functions
(also for differences of lower $n$-point functions at different
momenta) are usually dominated by the low momentum modes with
$q^2\approx k^2$. This motivates the use of SRPT rather
than standard perturbation theory for this purpose. Conversely,
the use of (approximated) full propagators and vertices as
determined by the solution of renormalization group or flow
equations may be advantageous for the estimate of some
dominant infrared effects (renormalons etc.),
which would only appear in higher order in standard
perturbation theory.

In summary, we have developed a systematic resummed
perturbation theory which is formulated in terms of full
propagators and vertices (eqs. (\ref{9}), (\ref{13}))
or approximations thereof (eqs. (\ref{17}),(\ref{21})).
The truncated version ressembles very closely standard perturbation
theory with classical action replaced by the truncated
effective action $\Gamma^{(tr)}$. Although we have
only sketched here some possible applications, we hope that we
have convinced some of the readers of this letter that SRPT can be
useful in practice for certain types of
problems in particle physics and statistical physics where standard
perturbative results become
problematic.

\section*{Figure captions}
\begin{description}
\item{Fig. 1:} Graphs contributing in two-loop order.
\item{Fig. 2:} One-loop contribution to the four-point function
from a six-point vertex.
\item{Fig. 3:} The two-loop contributions to the
four-point function resulting from contractions of the effective
six-point function in one-loop order.
\end{description}

\end{document}